\documentclass[a4paper]{jpconf}
\usepackage{graphicx}
\usepackage{units}
\begin{document}
\title{MiniBooNE-DM: a dark matter search in a proton beam dump}

\author{Alexis A. Aguilar-Arevalo
for the MiniBooNE-DM Collaboration}

\address{ Instituto de Ciencias Nucleares, Universidad Nacional Aut\'onoma de M\'exico, CDMX, 04510, M\'exico}

\ead{alexis@nucleares.unam.mx}

\begin{abstract}
In a dedicated run where protons from the Fermilab Booster were delivered directly to the steel beam dump of the Booster Neutrino Beamline (BNB), the MiniBooNE detector was used to search for the production of sub-GeV dark matter particles via vector-boson mediators.
The signal searched for was the elastic scattering of dark matter particles off nucleons in the detector mineral oil, with neutrinos being an irreducible background. 
A review of the experiment, its analysis methods, its results and future perspectives are summarized, demonstrating that beam dump experiments provide a novel and promising approach to dark matter searches.
This contribution was written for the TAUP 2017 proceedings.
\end{abstract}

\section{Introduction}

\hspace{0.4cm}
A large body of observations of gravitational phenomena \cite{pdg:2016} provide strong evidence for dark matter (DM), however, the lack of positive results from direct searches in the mass range from a few GeV to 10s of TeV \cite{direct} have motivated further extensions to the Standard Model (SM) featuring {\it Dark Sectors}, additional fields with no charges under the SM symmetries, which under appropriate assumptions are able to provide DM candidates in the sub-GeV mass range. There are only three renormalizable versions of the so called {\it portal models} connecting the dark sector to the SM that are constrained by the SM symmetries, differing on the type of mediator particle considered: the vector portal (vector mediator), the Higgs portal (scalar mediator),  and the neutrino portal (fermion mediator). Vector portal models are often considered among  the most viable for thermal sub-GeV DM. Accelerator beam dump experiments \cite{ds-fixedtarget} are powerful probes to explore this theoreically well motivated idea.\\

After completing its original physics program on neutrino oscillations and cross section measurements in 2012,
the MiniBooNE (MB) experiment at Fermilab conducted a special run in 2013-2014 with enhanced sensitivity to sub-GeV DM, by suppressing the neutrino production in the BNB. The results of this search \cite{aguilar:2017} demonstrated the unique capabilities of a neutrino experiment to search for DM.

\section{The BNB and the MiniBooNE Detector}

\hspace{0.4cm}
The standard neutrino-mode configuration of the BNB delivers 8 GeV protons to a 1.75 interaction length Be target in pulses of (3-5)$\times10^{12}$ protons, 1.6~$\mu$s in duration, producing a large flux of charged mesons, predominantly $\pi$'s. A pulsed toroidal magnetic field of $1.5$~T (magnetic horn) focuses the charged mesons along a 1 m-diameter, 50 m-long decay tunnel that ends in a steel beam dump. This creates a large neutrino flux at the detector location, 541 m from the target. Switching the polarity of the horn selects a beam of neutrinos or antineutrinos.\\

The MB detector contains 818 tons of mineral oil (CH$_2$) in a 12.2 m-diameter spherical tank. An inner wall at 550~cm radius is lined with 1280 photomultiplier tubes (PMTs) that view the main tank volume. An optically separate veto region 35~cm-thick is viewed by 240 PMTs arranged in pairs. PMT pulses with signal $>0.1$~photoelectron are digitized in a 19.2~$\mu$s window around the 1.6~$\mu$s BNB proton beam window. MB is primarily a Cherenkov detector but is also sensitive to the scintillation light emitted from trace fluors occurring in the oil as sub-Cherenkov particles traverse the medium.
It ran for more than 10 years in neutrino and antineutrino modes producing 27 publications on neutrino oscillations, cross section measurements and other physics, and it is very well understood. The DM search analysis heavily relies on the measurements of Charged Current Quasi-Elastic (CCQE) \cite{CCQE:2010} and Neutral Current Elastic (NCE) \cite{NCE:2015} scattering.

\section{Minimal kinetic mixing of a dark photon (MKMDP)}

\hspace{0.4cm}
The effective Lagrangian of a model minimally extending the SM with a single $U(1)_D$ ($D$ for {\it dark}) gauge boson, the {\it dark photon}, and a dark matter candidate $\chi$ is \cite{ds-fixedtarget}
\begin{eqnarray}
\nonumber
{\cal L}_{V,\chi} = |D_\mu\chi|^2 - m_\chi^2|\chi^2|
-\frac{1}{4}V_{\mu\nu}^2 + \frac{1}{2}m_V^2 V_{\mu}^2 
+ \epsilon V_{\mu\nu} F_{\mu\nu} + \dots \;\;,\\
\nonumber
D_\mu = \partial_\mu - i g_D V_\mu, \;\;\; g_D=\sqrt{4\pi\alpha_D} \;\;.
\hspace{2cm}
\end{eqnarray}
\noindent 
$F^{\mu\nu}$ is the SM electromagnetic field-strength tensor, and $V_{\mu\nu}$ is the corresponding object associated with the dark photon field $V_\mu$. The four model parameters are: the mediator particle mass $m_V$, the DM particle mass $m_\chi$, the kinetic mixing parameter $\epsilon$, and the dark photon interaction coupling strength $\alpha_D$. The presence of the mediator allows to adjust the DM annihilation cross section to match the observed relic density. A dark photon with a mass of $\mathcal{O}\sim 0.01-1$~GeV has been possited to explain the $(g-2)_\mu$ anomaly \cite{fayet:2007-pospelov:2009}.

\begin{figure}[t]
\centering
\scalebox{0.50}{\includegraphics{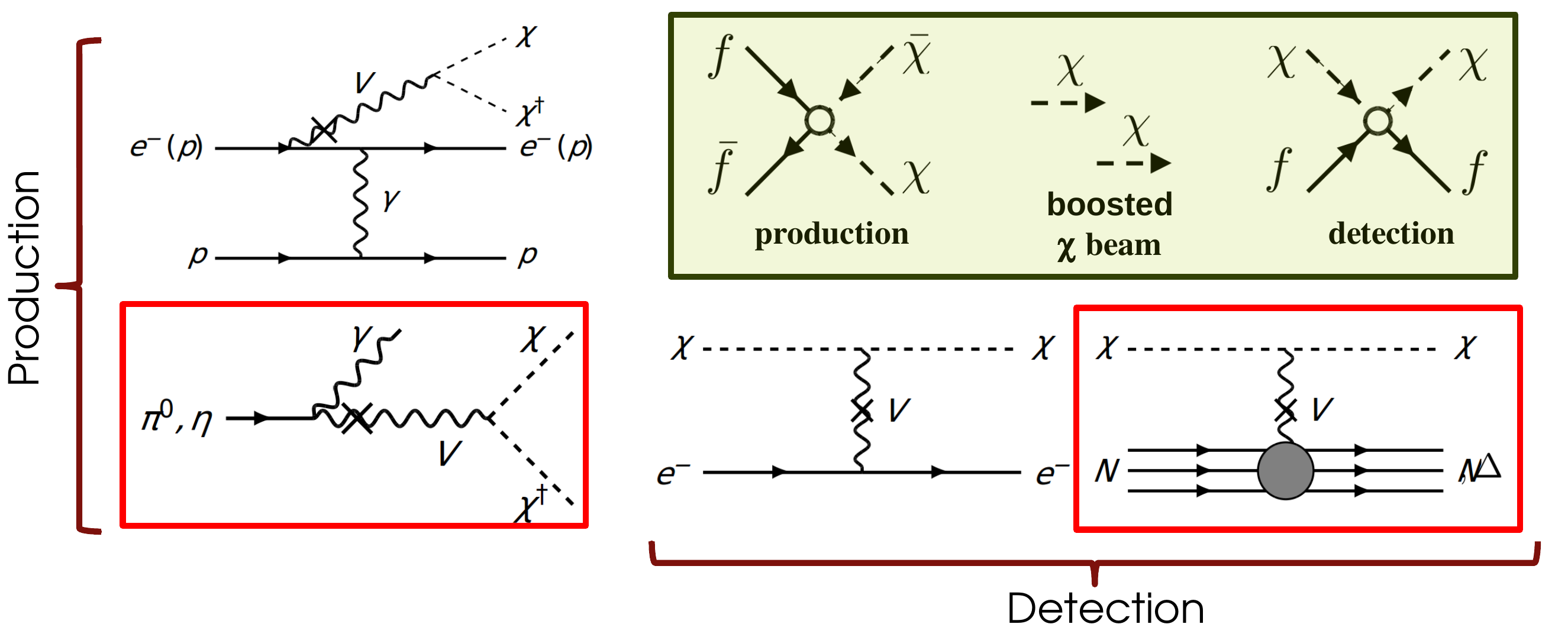}}
\caption{\small DM production is dominated by proton bremsstrahlung and decay of secondary $\pi^0$'s and $\eta$'s. Detection is expected through scattering off nucleons or electrons. The reactions most relevant to this search are those in the red boxes.}
\label{fig:prodet}
\end{figure}

\section{Production and detection of the dark matter beam}

\hspace{0.4cm}
DM can be produced in high-energy collisions of SM particles ({\it eg.} protons), and detected by its elastic scattering off nuclei in a large and sensitive detector via the mechanisms shown in Figure~\ref{fig:prodet}. The MKMDP model predicts that, when $m_V>2m_\chi$ ({\it invisible decay} of the $V$, which will be considered in what follows) the event rate scales as the combination $\epsilon^4\alpha_D$. \\

A large irreducible background to the DM search comes from neutrino interactions in the detector which therefore must be suppressed. The high-Z material of a beam dump will either absorb or stop charged mesons before they decay, significantly reducing the neutrino flux, compared to a standard neutrnio production target.

\section{Dark Matter search in off-target mode}

\begin{figure}[t]
\centering
\scalebox{0.64}{\includegraphics{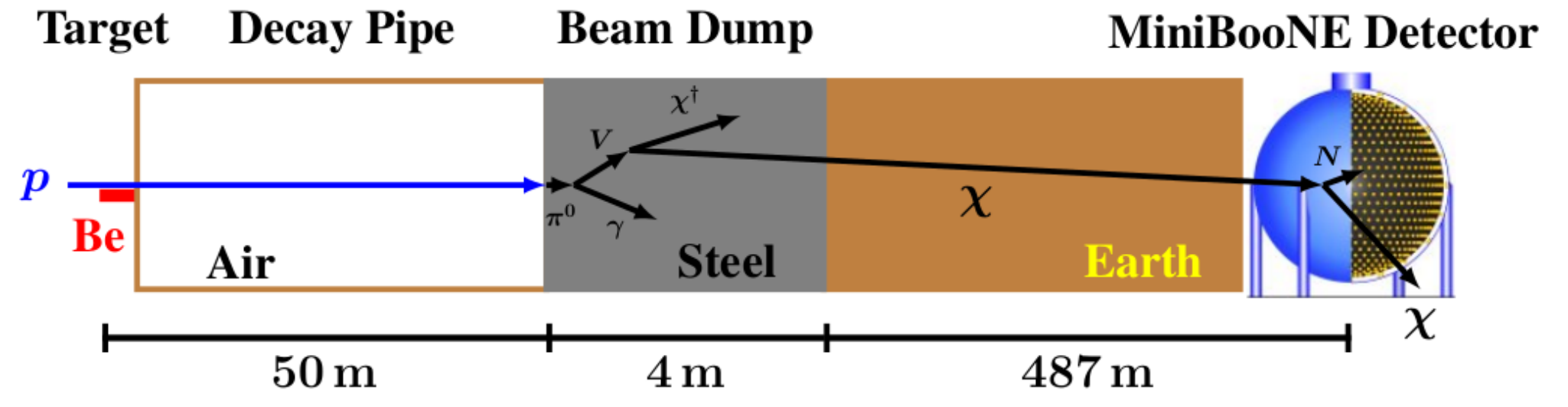}}
\caption{\small Schematic of the experimental setup in off-target mode. Taken from \cite{aguilar:2017}.}
\label{fig:schematic}
\end{figure}

\hspace{0.4cm}
For the DM search, the beam line was configured in ``off-target" mode, as shown in figure \ref{fig:schematic}. The protons were steered off the Be production target, keeping the magnetic horn powered off, and sent directly into the steel beam dump. At the the detector, $490$~m downstream of the beam dump, the neutrino flux is reduced by a factor of $\sim 30$ compared to neutrino-mode, while the CCQE event rate is $\sim$ 50 times smaller. A total of $1.86\times10^{20}$~POT (protons on target) were collected over the 9 months between November 2013 and  September 2014. \\

The selection of DM candidate events in the off-target data followed that of the MB antineutrino NCE analysis \cite{NCE:2015}, with the addition of a $35<T_n<600$~MeV cut on the reconstructed nucleon kinetic energy. The selection cuts require that one time-cluster of hits occurs in coincidence with the beam window, and have time and spatial distributions consistent with a single nucleon and no pions. A total of $1465\pm38$ events in the off-target data pass this selection.
Three additional ``constraint" samples were used to provide an improved estimate of beam-related backgrounds. A sample of CCQE events selected from the off-target data using the cuts developed for the CCQE cross-section measurement \cite{CCQE:2010}, was used to constrain the off-target neutrino flux, which gives rise to an irreducible background of neutrino-induced NCE events. Additionally, since the NCE and CCQE cross sections are not known independently of MB data, two other large samples with the same NCE and CCQE cuts as for the off-target data were selected from previously collected neutrino-mode data. \\

A total of $1579 \pm 529$ (34\% uncertainty) background events were expected before any constraints were applied: 697 events from beam-unrelated backgrounds (measured with an out-of-beam random trigger), 775 events from detector beam-related backgrounds (dominated by NCE neutrino interactions in the detector), and 107 events from beam-related ``dirt" backgrounds (neutrino-induced neutrons created outside the detector entering the main volume). A combined background-only fit to the three constraint samples and the off-target sample gives a constrained background prediction of $1548 \pm 198$ events (13\% uncertainty). The error estimates considered correlations between the different samples and energy bins. \\

Kinematic distributions for the nucleons in $\chi N$ scattering events for different values of $\epsilon^4 \alpha_D$, $m_V$, and $m_\chi$, were generated using the BdNMC DM simulation \cite{dmgen:2017}, and then passed on to the MB detector simulation. An effective efficiency calculated from the MB simulation was used to implement corrections for bound nucleons.

\begin{figure}[t]
\centering
\scalebox{0.57}{\includegraphics{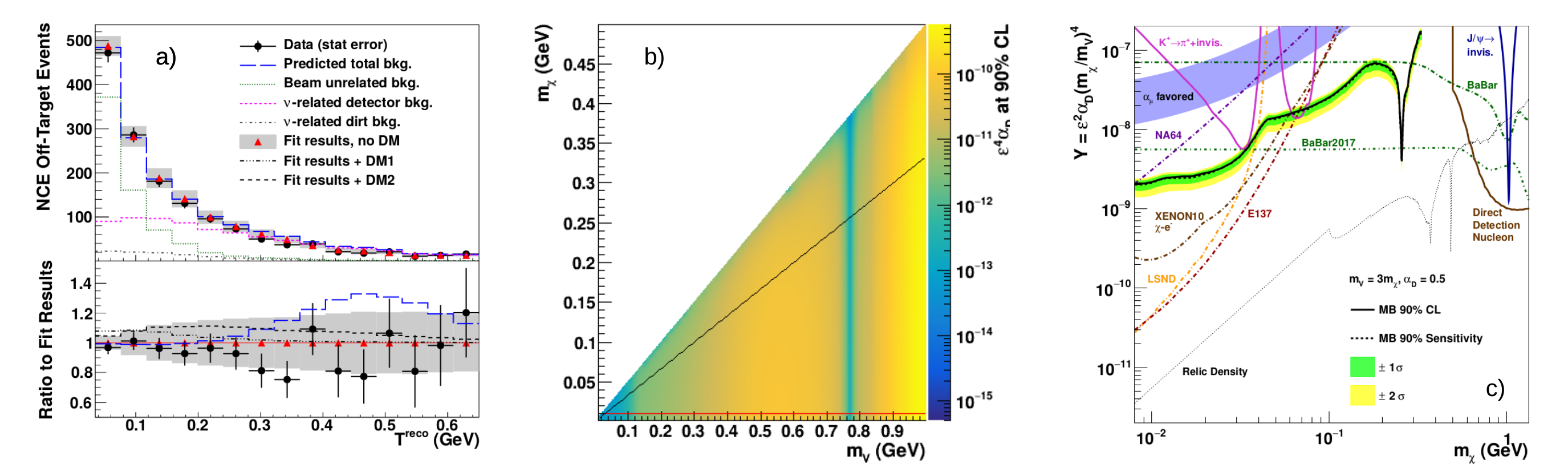}}
\caption{\small (a) Reconstructed nucleon kinetic energy distribution of DM candidates;
(b) the 90\% C.L. limit on $\epsilon^2\alpha_D$ for each point in the $m_\chi$ {\it vs} $m_V$ plane obtained with this search;
(c) the limit on the annihilation parameter $Y=\epsilon^2\alpha_D(m_\chi/m_V)^4$ that results from assuming the relation $m_V=3m_\chi$ and $\alpha_D=0.5$, compared to other results. The forthcoming results from BABAR \cite{babar:2017} are expected to be stronger, especially for smaller values of $\alpha_D$.}
\label{fig:results}
\end{figure}

\section{Results and future analyses}

\hspace{0.4cm}
A combined fit to the $Q^2$ distributions of the four samples, including the signal prediction on the NCE off-target sample, was used in a frequentist analysis to determine the 90\% C.L. limit on $\epsilon^4\alpha_D$ for given $m_V$ and $m_\chi$ (see figure \ref{fig:results}). The limits obtained from this search on the annihilation cross-section parameter $Y=\epsilon^2\alpha_D(m_\chi/m_V)^4$ fixing the parameters $m_V=3m_\chi$ and $\alpha_D=0.5$, compared to other experimental results, are shown in figure \ref{fig:results}-(c). Within the vector portal DM model and the chosen parameter constraints, the MB result excludes a vector mediator particle solution to the $(g-2)_\mu$ anomaly and sets stringent limits on DM with sub-GeV masses two orders of magnitude lower than attainable in nucleon direct detection experiments. \\

The collaboration is currently developing two new analyses: the ``$e$-DM" analysis, where the DM particles elastically scatter off electrons, and the ``{\it inelastic} $\pi^0$" analysis, where the signal is the resonant production of $\pi^0$'s in DM-nucleon interactions. These will increase the sensitivity to vector portal DM with the existing off-target data. An additional improvement expected to increase the sensitivity to DM masses $>70$~MeV will be to include the information of the timing structure of the Booster proton beam to look for characteristic out-of-time (intra bunch) events in the detector associated subluminal massive DM particles.

\section{Future perspectives}

\hspace{0.4cm}
This search has given valuable insight for the planning of future experiments. MB is a well characterized detector with over 10 years of data and a wealth of physics results; this provided a solid basis for understanding the backgrounds affecting the DM search. Being a large detector, the ability to tag incoming particles with an efficient veto in order to reduce the events originating in the surrounding dirt was very important. The need to measure the beam-unrelated backgrounds as accurately as possible was identified early on, and carried out effectively. A realistic estimate of the experiment's sensitivity requires to include the modeling of nuclear effects on the DM interaction with nucleons. Finally, several constraint samples which may be correlated an be used to reduce systematic uncertainties to a manageable level. \\

Simulation studies show that the neutrino flux could be further reduced by more than one order of magnitude, compared to off-target mode, by replacing the BNB target assembly with a dedicated Fe or W beam dump, hence removing 50 m of air along the protons flight path. This non-trivial upgrade to the BNB would represent a broadening of the original goals of the Fermilab Short Baseline Neutrino (SBN) Program to include a sub-GeV DM search. A short run collecting $2\times10^{20}$~POT with the SBN near detector (SBND) sitting 110~m downstream from such a beam dump, could achieve an order of magnitude better sensitivity in the $e$-DM and  {\it inelastic} $\pi^0$ channels. An SBND with an improved beam dump would be most sensitive to leptophobic models, where the DM does not couple to electrons.
These ideas have been presented to the community \cite{darksectors:2016,cosmicvisions:2017} and a LOI has been recently submitted to the Fermilab PAC.

\section{Conclusions}

\hspace{0.4cm}
Low-cost sub-GeV DM searches with proton accelerators can be done using the infrastructure of existing neutrino experiments with small modifications that enhance their capabilities. The successful MB beam dump run has already provided valuable lessons for improved future searches. This search was motivated by the vector portal model, however, other low-mass dark matter models are also possible. A very sensitive search for sub-GeV DM can be performed using the SBN near detector at Fermilab collecting data over a short \nicefrac{1}{2} to 1 year run. This will require a modest investment for an improved beam dump, with an increased return in science, leveraging on the existing neutrino program at the laboratory.

\section*{Acknowledgements}

\hspace{0.4cm}
The Collaboration thanks Fermilab, the Department of Energy, the National Science Foundation, and Los Alamos National Laboratory for supporting the experiment.
Fermilab is operated by Fermi Research Alliance, LLC under Contract No. DE-AC02-07CH11359 with the United States Department of Energy, 
Special thanks go to the Fermilab Accelerator Divison for reconfiguring, operating, and understanding the off-target beam.
The author acknowledges the support of DGAPA-UNAM, PAPIIT grant No.~IN108917, and of CONACYT.

\end{document}